\documentclass[fleqn,usenatbib]{mnras}

\usepackage[T1]{fontenc}
\usepackage{ae,aecompl}

\usepackage{graphicx}   
\usepackage{amsmath}    
\usepackage{amssymb}    
\usepackage{cuted}

\title[Thermal X-ray emission from magnetized WDs]{Thermal X-ray emission from massive, fast rotating, highly magnetized white dwarfs}

\author[D. L. C\'aceres et al.]{
D. L. C\'aceres,$^{1,2}$\thanks{diegoleonardocu@gmail.com}
S. M. de Carvalho,$^{3}$\thanks{sheyse.martins@icra.it}
J. G. Coelho,$^{4}$\thanks{jaziel.coelho@inpe.br}
R. C. R. de Lima,$^{2,5}$\thanks{rafael.camargo@icranet.org}\newauthor
Jorge A. Rueda,$^{1,2,6}$\thanks{jorge.rueda@icra.it}
\\
$^{1}$Dipartimento di Fisica and ICRA, Sapienza Universit\`a di Roma, P.le Aldo Moro 5, I--00185 Rome, Italy\\
$^{2}$ICRANet, P.zza della Repubblica 10, I--65122 Pescara, Italy\\
$^{3}$Universidade Federal do Tocantins, Campus Aragua\'ina, Av. Paraguai 77814-970 Aragua\'ina, TO, Brazil
\\
$^{4}$Div. Astrof\'isica, Instituto Nacional de Pesquisas Espaciais, Av. dos Astronautas 1758, 12227--010 S\~ao Jos\'e dos Campos, SP, Brazil\\
$^{5}$Universidade do Estado de Santa Catarina, Av. Madre Benvenuta, 2007 Itacorubi, 88.035--901, Florian\'opolis, Brazil\\
$^{6}$ICRANet-Rio, Centro Brasileiro de Pesquisas F\'isicas, Rua Dr. Xavier Sigaud 150, 2290--180 Rio de Janeiro, Brazil
}

\date{Accepted XXX. Received YYY; in original form ZZZ}

\pubyear{2016}

\begin{document}
\label{firstpage}
\pagerange{\pageref{firstpage}--\pageref{lastpage}}
\maketitle

\begin{abstract}
There is solid observational evidence on the existence of massive, $M\sim 1~M_\odot$, highly magnetized white dwarfs (WDs) with surface magnetic fields up to $B\sim 10^9$~G. We show that, if in addition to these features, the star is fast rotating, it can become a rotation-powered pulsar-like WD and emit detectable high-energy radiation. We infer the values of the structure parameters (mass, radius, moment of inertia), magnetic field, rotation period and spin-down rates of a WD pulsar death-line. We show that WDs above the death-line emit blackbody radiation in the soft X-ray band via the magnetic polar cap heating by back flowing pair-created particle bombardment and discuss as an example the X-ray emission of soft gamma-repeaters and anomalous X-ray pulsars within the WD model.
\end{abstract}




\section{Introduction}\label{sec:1}

The increasing data from observational campaigns leave no room for doubts on the existence of massive ($M\sim 1~M_\odot$) white  dwarfs (WDs) with  magnetic fields comprised in the range $B=10^6$--$10^{9}$~G \citep{2009A&A...506.1341K,2013MNRAS.429.2934K,2015MNRAS.446.4078K}. It has been recently shown that massive, highly magnetized WDs, could be formed by mergers of double WDs \citep{2012ApJ...749...25G}. The fact that WDs produced in mergers, besides being massive and highly magnetized, can be also fast rotators with periods $P\sim 10$~s, was used in \citet{2013ApJ...772L..24R} to show that they could be the WDs postulated in \citet{2012PASJ...64...56M} to describe the observational properties of soft gamma repeaters (SGRs) and anomalous X-Ray pulsars (AXPs), in alternative to the ``magnetar'' model \citep{1992ApJ...392L...9D,1995MNRAS.275..255T}. In the WD model the observed X-ray luminosity of SGRs/AXPs is explained via the loss of rotational energy of the fast rotating WD.\footnote{It remains open the case of 1E 161348--5055, the central compact object in the supernova remnant RCW 103, to be confirmed as a new SGR/AXP \citep[see, e.g.,][]{2016ApJ...828L..13R,2016MNRAS.463.2394D}. The observed luminosity and lightcurve periodicity with $P=6.67$~h, if confirmed to be due to the rotation period, together with the upper limit of to the possible spindown rate $|\dot{P}| < 7\times 10^{-10}$, rule out the rotational energy of either a neutron star or a WD as the possible source of energy.} The WD gravitational stability imposes a lower bound to the rotation period $P\approx 0.5$~s, in agreement with the minimum measured rotation period of SGRs/AXPs, $P\sim 2$~s \citep{2013ApJ...762..117B}. On the other hand, the surface area and temperature of the emitting region inferred from the available infrared, optical, and ultraviolet data of SGR/AXPs (i.e. for SGR 0418+5729, J1822.3--1606, 1E 2259+586 and 4U 0142+61), were shown to be consistent with the values expected from WDs \citep{2013A&A...555A.151B,2013ApJ...772L..24R}. The similarities of these WDs with ordinary, rotation-powered pulsars, imply that similar radiation mechanisms are expected to be at work in their magnetosphere. Indeed, the loss of rotational energy of the WD, owing to magnetic breaking, is sufficient to explain the X-ray luminosity observed in SGRs/AXPs, and the inferred magnetic field from the observed spindown rates, $B\sim 10^9$~G, agree with the aforementioned observed values in galactic WDs \citep{2012PASJ...64...56M,2014PASJ...66...14C}.

Following this line, it was advanced in \citet{2013ApJ...772L..24R} that the blackbody observed in the soft X-rays of SGRs/AXPs, with observationally inferred radii $R_{\rm bb}\sim 1$~km and temperatures $T_{\rm bb}\sim 10^6$~K, could be due to a known phenomenon expected to occur in pulsars, namely the magnetospheric currents flowing back towards the WD, heating up the magnetic polar caps creating surface hot spots \citep[see, e.g.,][]{1988SvAL...14..258U,1993ApJ...410..761U}. The aim of this work is to estimate this magnetospheric process for massive, highly magnetized, fast rotating WDs, exploiting the full analogy with pulsars. This calculation is interesting by its own and becomes of observational relevance in view of the latest results by \citet{2016Natur.537..374M} which point to the observational evidence of pulsar behavior of a magnetized WD. It is interesting that, as in the well-known case of AE Aquari, also this WD belongs to a binary system. Pulsar behavior can be observed from WDs in binaries when they can be considered approximately as isolated objects as in the case of detached binaries or in binaries in a propeller phase. Bearing this in mind, we focus in this work on the observable emission from isolated magnetized WDs in the X-rays. In order to exemplify the mechanism with appealing numbers, we apply it to the case of the WD model of SGRs and AXPs. Specifically, we evaluate the decay rate of curvature radiation photons in $e^-e^+$ pairs, and the subsequent backward flow of pair-produced particles that bombards and heats-up the magnetosphere polar caps, producing the observable thermal radiation. We compute in section \ref{sec:2} the condition for $e^-e^+$ pair creation within the inner gap model. In section \ref{sec:3} we calculate the expected thermal luminosity and infer the values of mass, radius, magnetic field, potential drop that ensure that the polar cap thermal emission explains the observed blackbody in the soft X-ray spectrum of SGRs/AXPs. The cases of 1E 2259+586 and 4U 0142+61 are analyzed as specific examples. In section \ref{sec:4} we simulate the observed X-ray flux from this spotty emission and compute the expected pulsed fraction which we compare with the observed values in SGRs and AXPs. We outline the conclusions in section \ref{sec:5}.

\section{WD magnetosphere}\label{sec:2}

Rotating, highly magnetized WDs can develop a magnetosphere analog to the one of pulsars. A corotating magnetosphere \citep{1947PhRv...72..632D,1949MNRAS.109..462F,1962JPSJS..17A.187G,1967ApJ...147..220F} is enforced up to a maximum distance given by the so-called light cylinder, $R_{\rm lc}=c/\Omega=c P/(2\pi)$, where $c$ is the speed of light and $\Omega$ is the angular velocity of the star, since corotation at larger distances would imply superluminal velocities for the  magnetospheric particles. For an axisymmetric star with aligned magnetic moment and rotation axes, the local density of charged plasma within the corotating magnetosphere is \citep{1969ApJ...157..869G}
\begin{equation}\label{eq:GJ}
\rho_{\rm GJ}=-\frac{\mathbf{\Omega}\cdot\mathbf{B}}{2\pi c}\frac{1}{1-(\Omega r_\bot/c)^2},
\end{equation}
where $r_{\bot}=r\sin\theta$ with $\theta$ the polar angle.

The last $B$-field line closing within the corotating magnetosphere can be easily located from the $B$-field lines equation for a magnetic dipole $r/\sin^2\theta=$ constant $=R_{\rm lc}$, and is located at an angle $\theta_{\rm pc} = \arcsin(\sqrt{R/R_{\rm lc}})\approx \sqrt{R/R_{\rm lc}}=\sqrt{R \Omega/c}=\sqrt{2\pi R/(c P)}$ from the star's pole, with $R$ the radius of the star. The $B$-field lines that originate in the region between $\theta=0$ and $\theta=\theta_{\rm pc}$ (referred to as \emph{magnetic polar caps}) cross the light cylinder, and are called ``open'' field lines. The size of the cap is given by the polar cap radius $R_{\rm pc}=R\,\theta_{\rm pc} \approx R\sqrt{2\pi R/(c P)}$. Clearly, by symmetry, there are two (antipodal) polar caps on the stellar surface from which the charged particles leave the star moving along the open field lines and escaping from the magnetosphere passing through the light cylinder.

Particle acceleration is possible in regions called \emph{vacuum gaps} where corotation cannot be enforced, i.e. where the density of charged particles is lower than the Goldreich-Julian value $\rho_{\rm GJ}$ given by Eq.~(\ref{eq:GJ}). For aligned(anti-aligned) rotation and magnetic axes we have $\rho_{\rm GJ}<0$ ($\rho_{\rm GJ}>0$), hence magnetosphere has to be supplied by electrons(ions) from the WD surface. This work is done by the existence of an electric field parallel to the magnetic field. Independently on whether $\mathbf{\Omega}\cdot\mathbf{B}$ is positive or negative, we assume that the condition of a particle injection density lower than $\rho_{\rm GJ}$ is fulfilled. In this \emph{inner gap} model the gaps are located just above the polar caps \citep{1975ApJ...196...51R} and the potential drop generated by the unipolar effect and that accelerates the electrons along the open $B$-field lines above the surface is
\begin{equation}\label{eq:Vgral}
\Delta V=\frac{B_s\Omega h^2}{c},
\end{equation}
where $h$ is the height of the vacuum gap and $B_s$ is the surface magnetic field, which does not necessarily coincides with the dipole field $B_p$.

The electrons (or positrons) accelerated through this potential and following the $B$-field lines will emit curvature photons whose energy depends on the $\gamma$-factor, $\gamma=e\Delta V/(m c^2)$, where $e$ and $m$ are the electron charge and mass, and on the $B$-field line curvature radius $r_c$, i.e. $\omega_c=\gamma^3 c/r_c$. Following \citet{1993ApJ...402..264C}, we adopt the constraint on the potential $\Delta V$ for pair production via $\gamma + B \to e^- + e^+$,
\begin{equation}\label{eq:eVpair}
\frac{1}{2}\left(\frac{e\Delta V}{m c^2}\right)^{3}\frac{\lambda}{r_c}\frac{h}{r_c}\frac{B_s}{B_q}\approx \frac{1}{15},
\end{equation}
or in terms of a condition on the value of the potential,
\begin{equation}\label{eq:eVpair2}
\Delta V\approx \left(\frac{2}{15}\right)^{2/7}\left( \frac{r_c}{\lambda} \right)^{4/7} \left( \frac{\lambda \Omega}{c} \right)^{1/7} \left( \frac{B_s}{B_q} \right)^{-1/7} \frac{m c^2}{e},
\end{equation}
where we have used Eq.~(\ref{eq:Vgral}), $\lambda=\hbar/(m c)$, $B_q\equiv m^2 c^3/(e\hbar)=4.4\times 10^{13}$~G, is the quantum electrodynamic field, with $\hbar$ the reduced Planck's constant.

For a magnetic dipole geometry, i.e. $B_s=B_d$ and $r_c=\sqrt{R c/\Omega}$, the potential drop $\Delta V$ cannot exceed the maximum potential  (i.e., for $h=h_{\rm max}=R_{\rm pc}/\sqrt{2}$), 
\begin{equation}\label{eq:Vmax}
\Delta V_{\rm max}=\frac{B_d\Omega^2 R^3}{2 c^2}.
\end{equation}

We are here interested in the possible magnetospheric mechanism of X-ray emission from magnetized WDs, thus we will consider the heating of the polar caps by the inward flux of pair-produced particles in the magnetosphere. These particles of opposite sign to the parallel electric field moves inward and deposit most of their kinetic energy on an area 
\begin{equation}\label{eq:AspotIG}
A_{\rm spot}=f A_{\rm pc},
\end{equation}
i.e. a fraction $f\leq 1$ of the polar cap area, $A_{\rm pc}=\pi R_{\rm pc}^2$. The temperature $T_{\rm spot}$ of this surface hot spot can be estimated from the condition that it re-radiates efficiently the deposit kinetic energy, as follows. The rate of particles flowing to the polar cap is $\dot{N}=J A_{\rm pc}/e$, where $J=\eta \rho_{\rm GJ} c$ is the current density in the gap, and $\eta <1$ a parameter that accounts for the reduction of the particle density in the gap with respect to the Goldreich-Julian value (\citealp{1977ApJ...214..598C} used $\eta= 1$ for order-of-magnitude estimates). In this model the filling factor $f$ is not theoretically constrained and it has been estimated from pulsar's observations in X-rays that it can be much smaller than unity \citep{1977ApJ...214..598C}. The condition that the hot spot luminosity equals the deposited kinetic energy rate reads
\begin{equation}\label{eq:Tspot1}
A_{\rm spot}\sigma T_{\rm spot}^{4}= e \Delta V \dot{N} = J A_{\rm pc} \Delta V = \eta \rho_{\rm GJ}(R) c A_{\rm pc} \Delta V,
\end{equation}
where $\sigma$ is the Stefan-Boltzmann constant. From Eqs.~(\ref{eq:GJ}), (\ref{eq:AspotIG}) and (\ref{eq:Tspot1}) we obtain the spot temperature
\begin{equation}\label{eq:TspotIG}
T_{\rm spot}=\left(\eta \frac{B_d\Delta V}{\sigma f P} \right)^{1/4}.
\end{equation}
It is worth to mention that in the above estimate we have assumed a full efficiency in the conversion from the deposited kinetic energy to the hot spot emission. This assumption is accurate if the heating source namely the energy deposition occurs not too deep under the star's surface and it is not conducted away to larger regions being mainly re-radiated from the surface area filled by the penetrating particles \citep{1980ApJ...235..576C}. In Appendix~\ref{app:A} we estimate the cooling and heating characteristic times and the heating and re-radiation efficiency. For the densities and temperatures of interest here we show that the polar cap surface re-radiates efficiently most of the kinetic energy deposited by the particle influx validating our assumption.

\section{Specific examples}\label{sec:3}

As in our previous analyzes \citep{2012PASJ...64...56M,2013A&A...555A.151B,2013ApJ...772L..24R,2014PASJ...66...14C}, we use the traditional dipole formula to get an estimate for the WD dipole magnetic field, i.e.:
\begin{equation}\label{eq:Bdipole}
B_d=\left(\frac{3 c^3}{8 \pi^2} \frac{I}{R^6} P \dot{P} \right)^{1/2},
\end{equation}
where $I$ is the moment of inertia of the star, $\dot{P}\equiv dP/dt$ is the first time derivative of the rotation period (spindown rate), and an inclination of $\pi/2$ between the magnetic dipole and the rotation axis, has been adopted. It is worth to recall that the estimate of the $B$-field by \eqref{eq:Bdipole} is not necessarily in contrast, from the quantitative point of view, with an estimate using an aligned field but introducing a breaking from the particles escaping from the magnetosphere, since also in this case a quantitatively and qualitatively similar energy loss is obtained.

For a given rotation period $P$, the WD structure parameters such as mass $M$, radius $R$, and moment of inertia $I$ are bounded from below and above if the stability of the WD is requested \citep{2013A&A...555A.151B}. From those bounds, we established there lower and upper bounds for the field $B_d$ of the WD.

\subsection{1E 2259+586}\label{sec:3.1}

We apply the above theoretical framework to a specific source, AXP 1E 2259+586. This source, with a rotation period $P = 6.98$~s \citep{fahlman81} and a spindown rate $\dot{P}= 4.8\times 10^{-13}$ \citep{davies90}, has a historical importance since \cite{1990ApJ...365L...9P} first pointed out the possibility of this object being a WD. This object produced a major outburst in 2002 \citep{2003ApJ...588L..93K, 2004ApJ...605..378W}, in which the pulsed and persistent fluxes rose suddenly by a factor of $\geq 20$ and decayed on a timescale of months. Coincident with the X-ray brightening, the pulsar suffered a large glitch of rotation frequency fractional change $4\times 10^{-6}$ \cite{2003ApJ...588L..93K, 2004ApJ...605..378W}. It is worth to recall that the observed temporal coincidence of glitch/bursting activity, as first pointed out by \cite{1994ApJ...427..984U} in the case of 1E 2259+586, and then extended in \cite{2012PASJ...64...56M,2013A&A...555A.151B}, can be explained as due to the release of the rotational energy, gained in a starquake occurring in a total or partially crystallized WD. Since we are interested in the quiescent behavior, we will not consider this interesting topic here. Therefore, only X-ray data prior to this outburst event will be used in this work \citep{2008ApJ...686..520Z}.

The soft X-ray spectrum of 1E 2259+586 is well fitted by a blackbody plus a power law model. The blackbody is characterized by a temperature $k T_{\rm bb}\approx 0.37$~keV ($T_{\rm bb}\approx 4.3\times 10^6$~K) and emitting surface are $A_{\rm bb}\approx 1.3\times 10^{12}$~cm$^2$ \citep{2008ApJ...686..520Z}. These values of temperature and radius are inconsistent (too high and too small, respectively) with an explanation based on the cooling of a hot WD, and therefore such a soft X-ray emission must be explained from a spotty surface due to magnetospheric processes, as the one explored in this work.

The stability of the WD for such a rotation period constrains the WD radius to the range $R\approx (1.04$--$4.76)\times 10^8$~cm. 
For example, in the case of a WD with radius $R\approx 10^8$~cm, the polar cap area is $A_{\rm pc}=6.6\times 10^{14}$~cm, hence using Eq.~(\ref{eq:AspotIG}) we have $f\approx 0.002$ and from Eq.~(\ref{eq:TspotIG}). The spot temperature $k\,T_{\rm spot}\approx 0.37$~keV can be obtained using $B_d\approx 6\times 10^9$~G from the dipole formula (\ref{eq:Bdipole}), a potential drop $\Delta V\approx 3.5\times 10^{11}$~Volts (lower than $\Delta V_{\rm max}\approx 5.4\times 10^{12}$~Volts), and using the typical value $\eta=1/2$ of the reduced particle density in the gap adopted in the literature. These parameters suggests a height of the gap, obtained with Eq.~(\ref{eq:Vgral}), $h\approx 0.11 R_{\rm pc}$. 

\begin{figure}
\centering
\includegraphics[width=\hsize,clip]{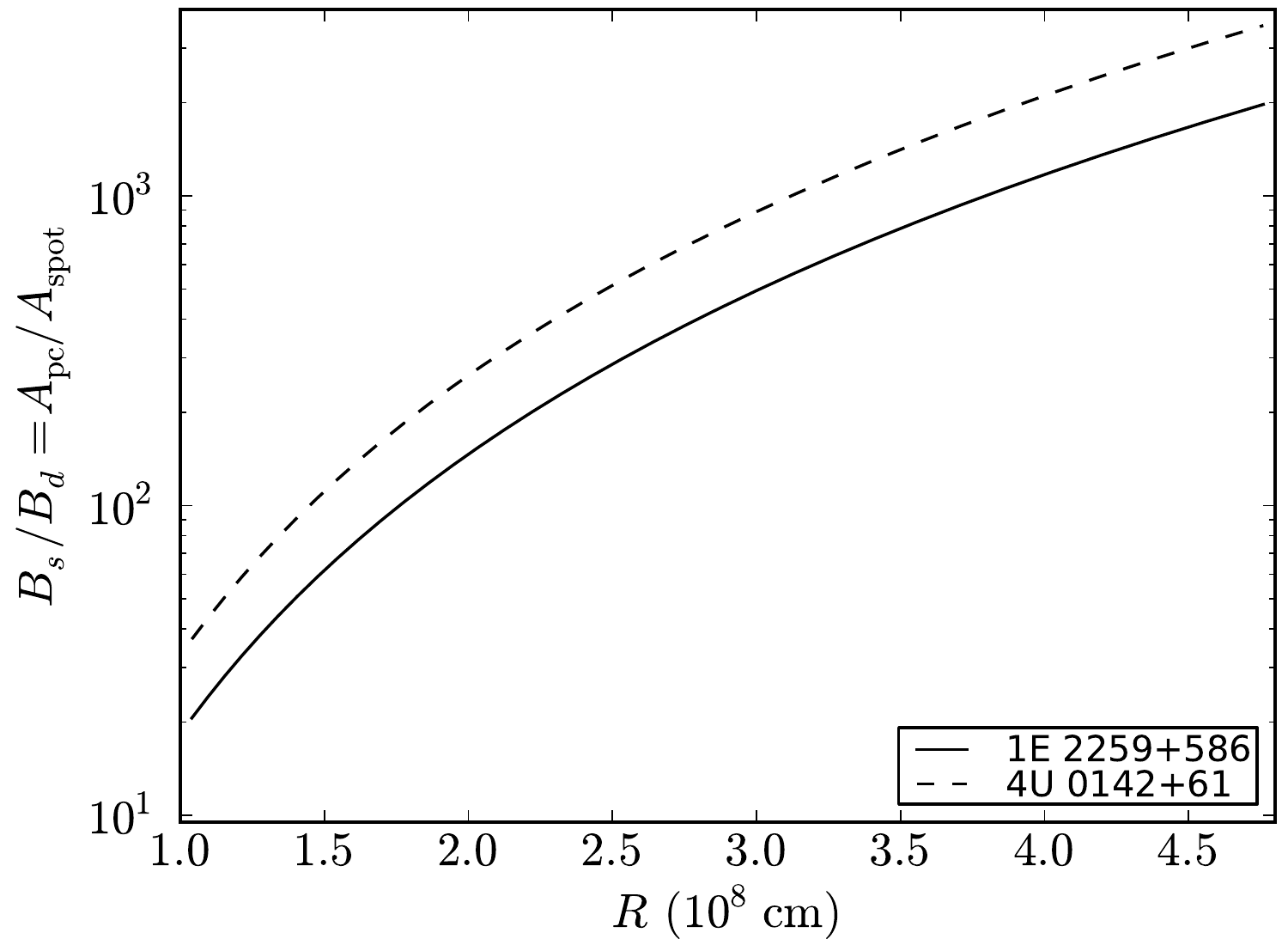}
\caption{Surface to dipole magnetic field ratio given by magnetic flux conservation (\ref{eq:BmIG}) for the AXPs 1E 2259+586 and 4U 0142+61.}\label{fig:figure1}
\end{figure}

The smallness of the filling factor, which appears to be not attributable to the value of $h$, could be explained by a multipolar magnetic field near the surface. It is interesting that the existence of complex multipolar magnetic field close to the WD surface is observationally supported \citep[see, e.g.,][]{2015SSRv..191..111F}. It is important to clarify that the above defined filling factor has only a physical meaning when, besides a strong non-dipolar surface field, the physical parameters of the star (magnetic field and rotational velocity) fulfill the requirement for the creation of electron-positron pairs in such a way that an avalanche of particles hits the surface. This is given by the request that the potential drop (\ref{eq:eVpair2}) does not exceed the maximum value (\ref{eq:Vmax}). For example, for the largest magnetic field measured in WDs, $B\sim 10^9$G, and WD radii $10^8$--$10^9$~cm, the maximum period that allows the avalanche of electron-positron pairs gives the range $P\sim 4$--$100$~s, values much shorter value than the typical rotation periods measured in most of magnetic WDs, $P\gtrsim 725$~s \citep[see, e.g.,][]{2015SSRv..191..111F}. It is interesting to note that the condition of $e^+e^-$ pair creation in the WD magnetosphere could explain the narrow range of observed rotation periods of SGRs/AXPs, $P\sim 2$--$12$~s. Such a local, strong non-dipolar field in the surface, diminishes the area bombarded by the incoming particles and, via magnetic flux conservation, the filling factor establishes the intensity of the multipolar magnetic field component as \citep[see, e.g.,][and references therein]{1999ApJ...515..337C,2000ApJ...541..351G,2002ApJ...577..909G}
\begin{equation}\label{eq:BmIG}
B_s = \frac{B_d}{f},
\end{equation}
which implies that, close to the surface, there could be small magnetic domains with magnetic field intensity as large as $10^{11}-10^{12}$~G (see Fig.~\ref{fig:figure1}).

\subsection{4U 0142+61}\label{sec:3.2}

We can repeat the above analysis for the case of 4U 0142+61. This source, with a rotation period $P\approx 8.69$~s, was first detected by \textit{Uhuru} \citep{1978ApJS...38..357F}. The measured period derivative of this source is $\dot{P}=2.03\times 10^{-12}$ \citep{2000Natur.408..689H}. The time-integrated X-ray spectrum of 4U 0142+61 is also described by a blackbody plus a power-law model. The blackbody component shows a temperature $k T_{\rm bb}=0.39$~keV ($T_{\rm bb}\approx 4.6\times 10^{6}$~K) and a surface area $A_{\rm bb}\approx 5.75\times 10^{11}$~cm$^2$ \citep{2005A&A...433.1079G}. As for the above case of 1E 2259+5726, such a blackbody cannot be explained from the cooling of a WD but instead from a magnetospheric hot spot created by the heating of the polar cap. 

For a WD radius $R=10^8$~cm and a magnetic field $B_d\approx 10^{10}$~G for a rotating dipole (\ref{eq:Bdipole}), we have a filling factor $f\approx 0.001$, a potential drop $\Delta V\approx 1.4\times 10^{11}$~Volts (smaller than $\Delta V_{\rm max}\approx 5.8\times 10^{12}$~Volts) and a gap height $h\approx 0.06 R_{\rm pc}$. Again the filling factor suggests the presence of a strong multipolar component as shown in Fig.~\ref{fig:figure1}.

\begin{figure*}
\centering
\includegraphics[width=0.5\hsize,clip]{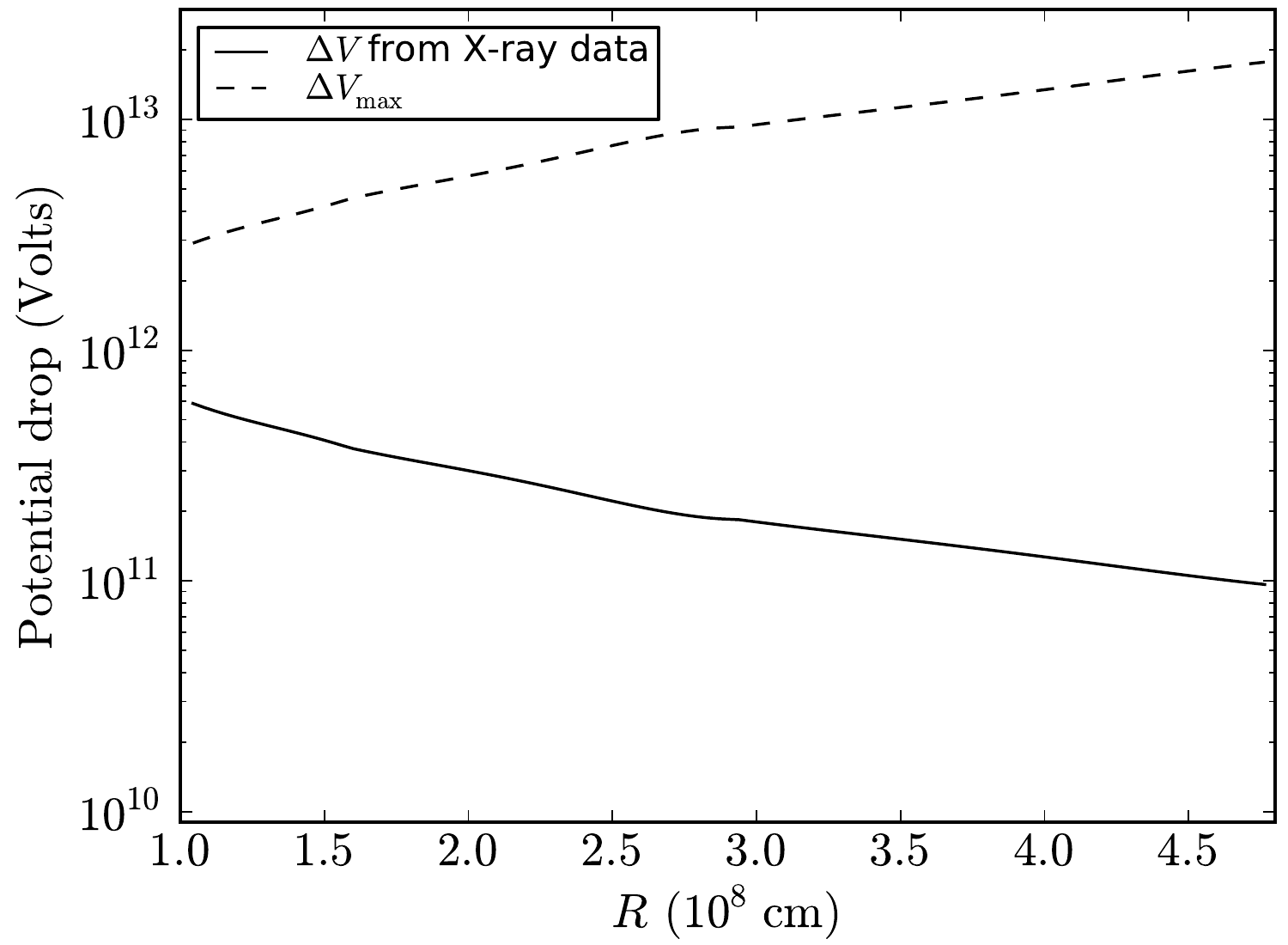}\includegraphics[width=0.5\hsize,clip]{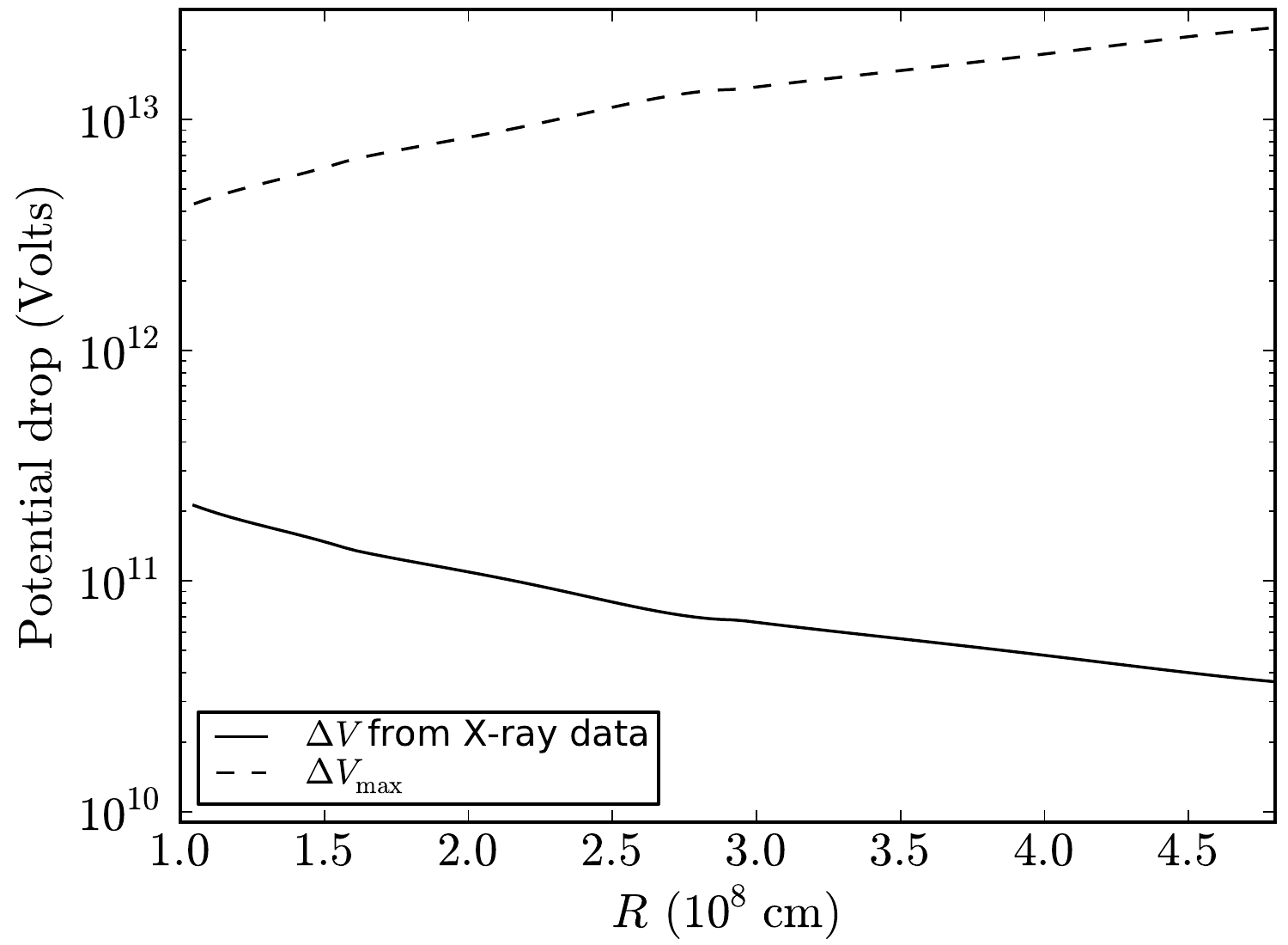}
\caption{WD polar gap potential drop $\Delta V$ inferred via Eq.~(\ref{eq:TspotIG}) using the blackbody observed in soft X-rays in 1E 2259+586 (left panel) and 4U 0142+61 (right panel). In this plot we check the potential drop developed in the WD polar gap does not exceed the maximum potential reachable $\Delta V_{\rm max}$ given by Eq.~(\ref{eq:Vmax}).}\label{fig:Vmax}
\end{figure*}

We show in Fig.~\ref{fig:Vmax} the potential drop inferred from Eq.~(\ref{eq:TspotIG}) using the X-rays blackbody data for the above two sources. We check that for all the possible stable WD configurations the potential drop satisfies the self-consistence condition $\Delta V < \Delta V_{\rm max}$, where the latter is given by Eq.~(\ref{eq:Vmax}).

\section{Flux Profiles and Pulsed Fraction}\label{sec:4}

We turn now to examine the properties of the flux emitted by such hot spots. Even if the gravitational field of a WD is not strong enough to cause appreciable general relativistic effects, for the sake of generality we compute the flux from the star taken into account the bending of light.
%
We shall follow here the treatment in \citet{2013ApJ...768..147T} to calculate the observed flux, which allows to treat circular spots of arbitrary finite size and arbitrarily located in the star surface. The mass and radius of the star are denoted by $M$ and $R$, and the outer spacetime is described by the Schwarzschild metric, i.e. we shall neglect at first approximation the effects of rotation. Let be $(r,\theta,\phi)$ the spherical coordinate system centered on the star and the line of sight (LOS) the polar axis.
\begin{figure}
\centering
\includegraphics[width=0.6\hsize,clip]{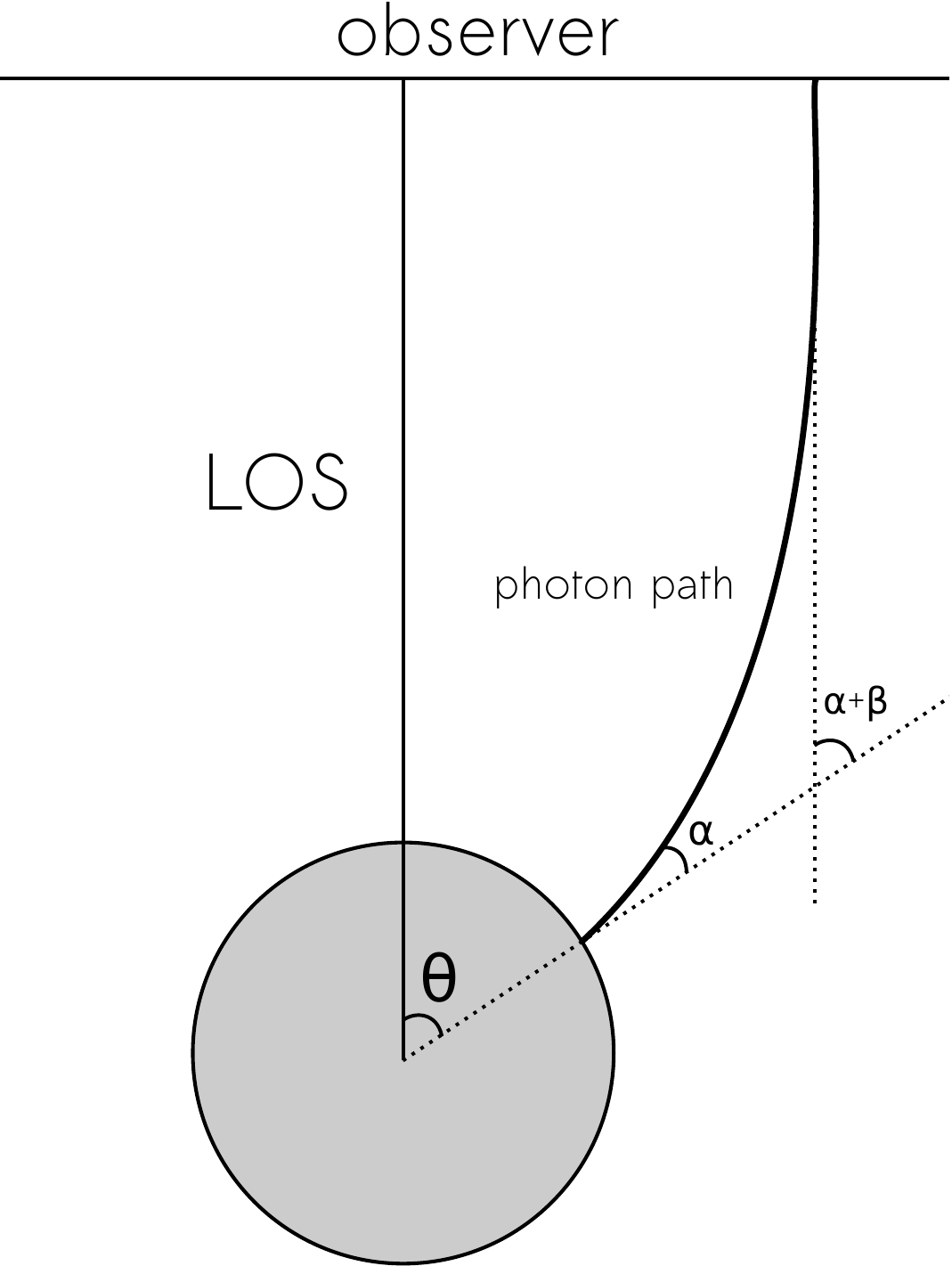}
\caption{View of the photon trajectory and angles $\theta$, $\alpha$, and $\beta$.}\label{photonpath}
\end{figure}

We consider an observer at $r\rightarrow \infty$ and a photon that arises from the star surface at $dS=R^2 \sin\theta d\theta d\phi$ making an angle $\alpha$ with the local surface normal, where $ 0 \leq \alpha \leq \pi/2$. The photon path is then bended by an additional angle $\beta$ owing to the spacetime curvature, reaching the observer with an angle $\psi=\alpha+\beta$. Since we have chosen the polar axis aligned with the LOS, it is easy to see that $\psi=\theta$ (see Fig.~\ref{photonpath}). \citet{2002ApJ...566L..85B} showed that a simple approximate formula can be used to relate the emission angle $\alpha$ to the final angle $\theta$:
\begin{equation}
    1 -\cos\alpha = (1-\cos\theta)\left(1-\frac{R_s}{R}\right),
\end{equation}
where $R_s=2GM/c^2$ is the Schwarzschild radius and, as usual, $G$ denotes the gravitational constant.

For an emission with a local Planck spectrum, the intensity is given by a blackbody of temperature $T$, $B_\nu(T)$, where $\nu$ is the photon frequency. The flux is proportional to the visible area of the emitting region ($S_V$) plus a relativistic correction proportional to the surface, given by the equation
\begin{align}\label{FluxTurolla}
    F_\nu &= \left(1-\frac{Rs}{R}\right) B\nu(T) \int_{S_V} \cos\alpha \frac{dcos\alpha}{d(\cos\theta)}ds \nonumber \\
    &=\left(1-\frac{Rs}{R}\right)^2 B_\nu (T)(I_p + I_s),
\end{align}
where
\begin{equation}
 I_p = \int_{S_V} \cos\theta \sin\theta d\theta d\phi,\quad I_s = \int_{S_V} \sin\theta d\theta d\phi .
\end{equation}
In polar coordinates, the circular spot has its center at $\theta_0$ and a semi-aperture $\theta_c$. The spot is bounded by the function $\phi_b(\theta)$, where $0\leq \phi_b \leq \pi$, and since we must consider just the star visible part, the spot must be also limited by a constant $\theta_F$. For a given bending angle $\beta$, the maximum $\theta_F$ is given by the maximum emission $\alpha$, i.e. $\alpha=\pi/2$. One can see that in a Newtonian gravity, where $\beta=0$, the maximum visible angle is $\theta_F=\pi/2$ which means half of the star is visible, while in a relativistic star, values $\theta_F>\pi/2$ are possible, as expected. Then
\begin{align}
    & I_p = 2 \int_{\theta_{min}}^{\theta_{max}} \cos\theta \sin\theta \phi_b(\theta) d\theta, \nonumber \\
    & I_s = 2 \int_{\theta_{min}}^{\theta_{max}} \sin\theta \phi_b(\theta) d\theta,
\end{align}
where $\theta_{min}$, $\theta_{max}$ are the limiting values to be determined for the spot considered.
\citet{2013ApJ...768..147T} showed how to solve these integrals and how to treat carefully the limiting angles. The $I_p$ and $I_s$ integrals can be then written as $I_{p,s}=I_{1,2}(\theta_{max})-I_{1,2}(\theta_{min})$ and we refer the reader to that work for the precise expressions.
 %
%
Finally, the flux (\ref{FluxTurolla}) is written as
\begin{equation}
    F_\nu = \left(1-\frac{Rs}{R}\right)^2 B_\nu(T) A_{\rm eff}(\theta_c,\theta_0) \ ,
\end{equation}
where $A_{\rm eff}$ is the effective area, given by
\begin{equation}
    A_{\rm eff}(\theta_c,\theta_0) = R^2 \left[ \frac{Rs}{R}I_s + \left(1-\frac{Rs}{R}\right)I_p\right] \ .
\end{equation}

The total flux produced by two antipodal spots, with semi-apertures $\theta_{c,i}$ and temperatures $T_i$ (i=1,2), can be calculated by adding each contribution, so we have
\begin{eqnarray}\label{Antipodalflux}
    F_\nu^{TOT} &=&  \left(1-\frac{Rs}{R}\right)^2 [B_\nu(T_1) A_{\rm eff}(\theta_{c,1},\theta_0) \nonumber\\
    & +&  B_\nu(T_2) A_{\rm eff}(\theta_{c,2},\theta_0 + \pi/2) ] \ .
\end{eqnarray}
Besides, the pulse profile in a given energy band $[\nu_1,\nu_2]$ for one spot is given by
\begin{equation}
    F(\nu_1,\nu_2)=\left(1-\frac{Rs}{R} \right)^2 A_{\rm eff}(\theta_c,\theta_0) \int_{\nu_1}^{\nu_2} B_\nu(T)d\nu \ .
\end{equation}

\begin{figure}
\centering
\includegraphics[width=\hsize,clip]{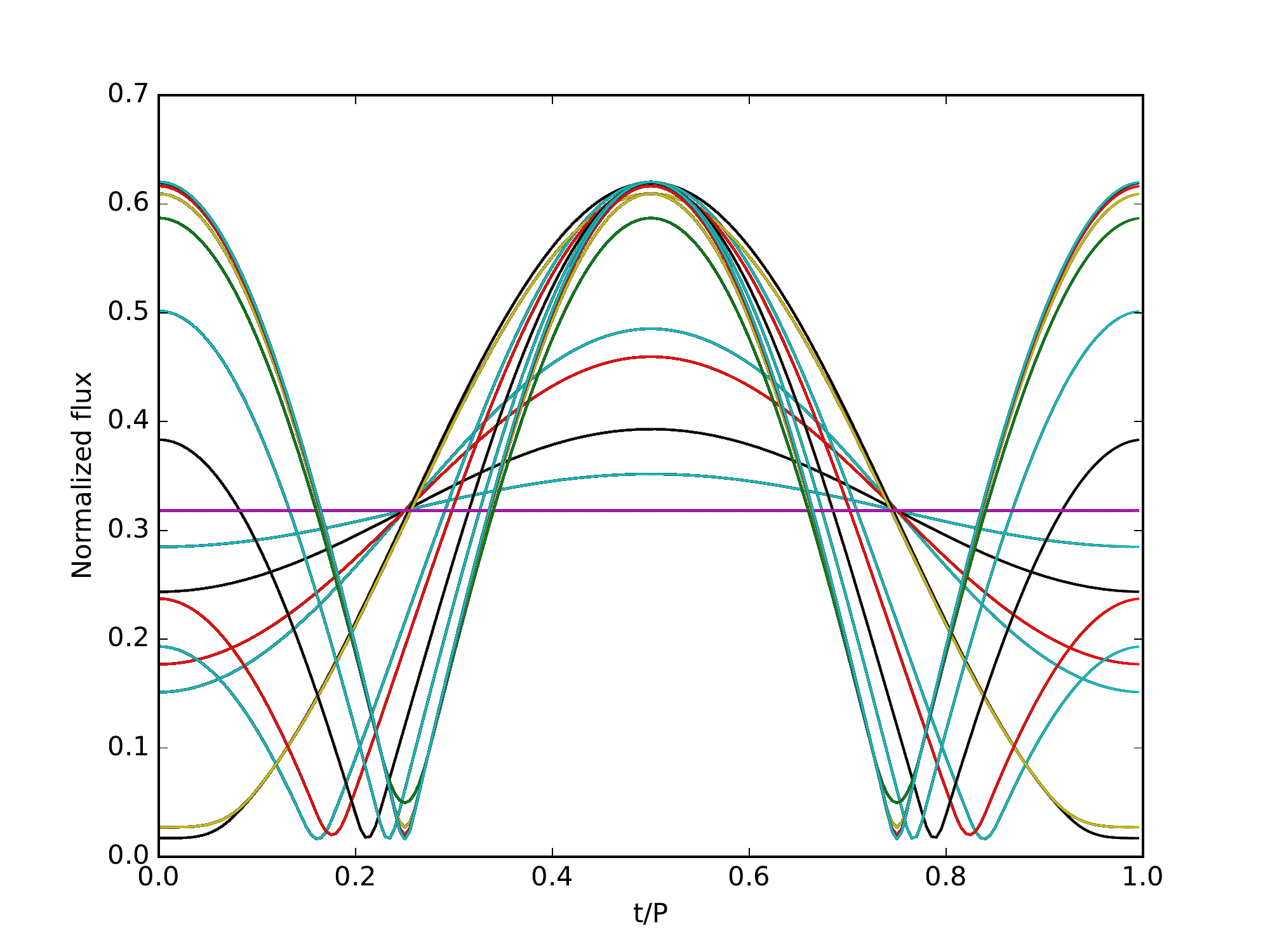}
\caption{Flux profiles for different configurations of antipodal spots as a function of the phase. The semi-aperture for all the lines is $\theta_c=3^{\circ}$. The WD parameters correspond to the ones of the WD of minimum radius adopted for AXP 1E 2259+586.}\label{fig:fluxExamp}
\end{figure}

The star rotates with a period $P$ (angular velocity $\Omega=2\pi/P$), so we consider $\hat{\mathbf r}$ the unit vector parallel to the rotating axis. It is useful to introduce the angles $\xi$, the angle between the LOS (unit vector $\hat{\mathbf l}$) and the rotation axis, and the angle $\chi$ between the spot axis (unit vector $\hat{\mathbf c}$) and the rotation axis, i.e., $\cos\xi=\hat{\mathbf r}\cdot\hat{\mathbf l}$ and $\cos\chi=\hat{\mathbf r}\cdot\hat{\mathbf c}$. As the star rotates the spot's center, $\theta_0$, changes. Let $\gamma(t)=\Omega t$ be the rotational phase, thus by geometrical reasoning we have
\begin{equation}
    \cos\theta_0(t)=\cos\xi\cos\chi-\sin\xi\sin\chi\cos\gamma(t) \ ,
\end{equation}
where it is indicated that $\xi$ and $\chi$ do not change in time. When the total flux (\ref{Antipodalflux}) is calculated for a given configuration $(\xi,\chi)$ in the whole period of time, the typical result is a pulsed flux with a maximum ($F_{\rm max}$) and a minimum flux ($F_{\rm min}$). As an example, we show in Fig.~\ref{fig:fluxExamp} flux profiles for different configurations of antipodal spots as a function of the phase for the WD of minimum radius in the case of AXP 1E 2259+586 used in Sec.~\ref{sec:3.1}.

\begin{figure*}
\centering
\includegraphics[width=0.5\hsize,clip]{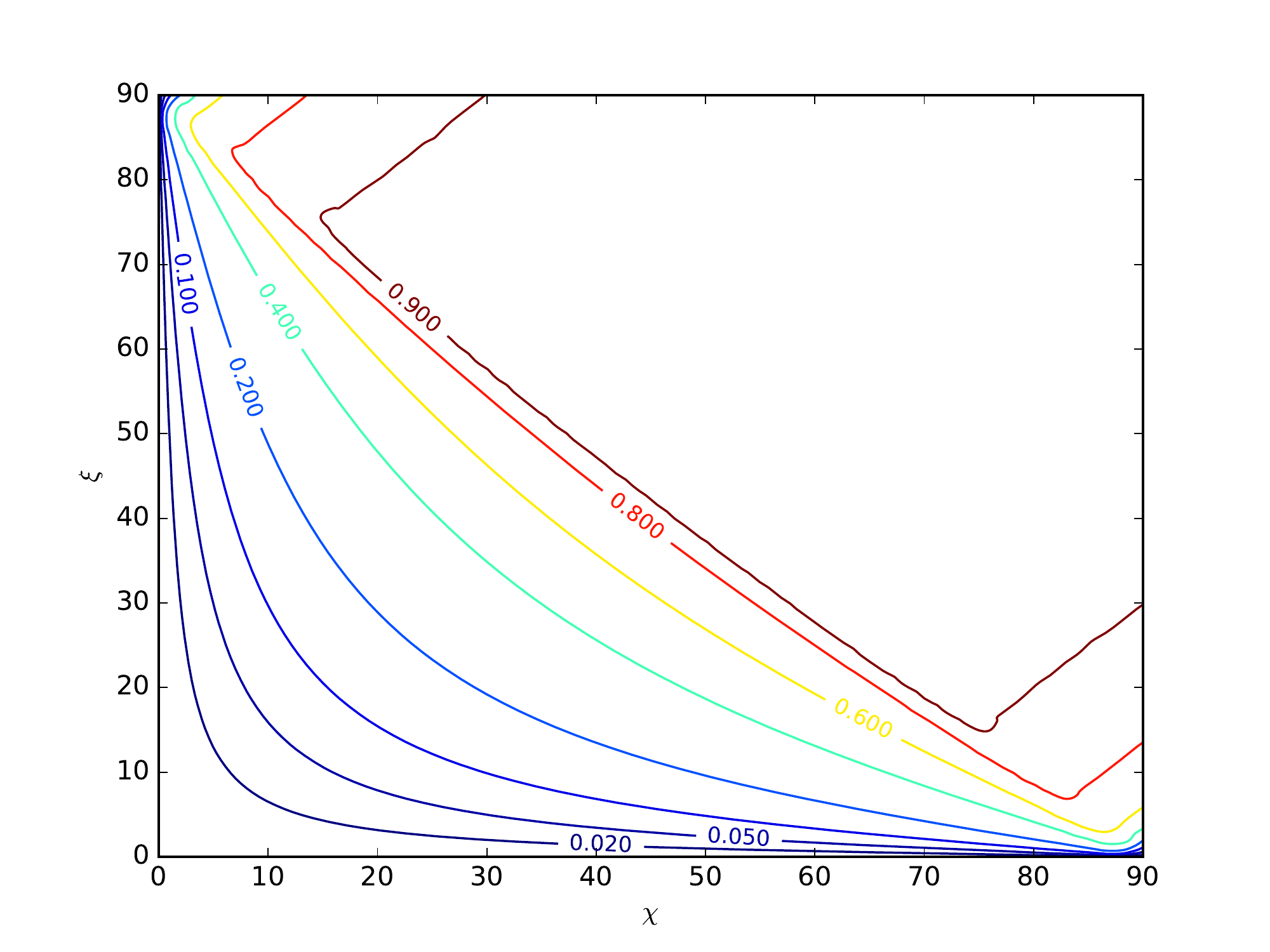}\includegraphics[width=0.5\hsize,clip]{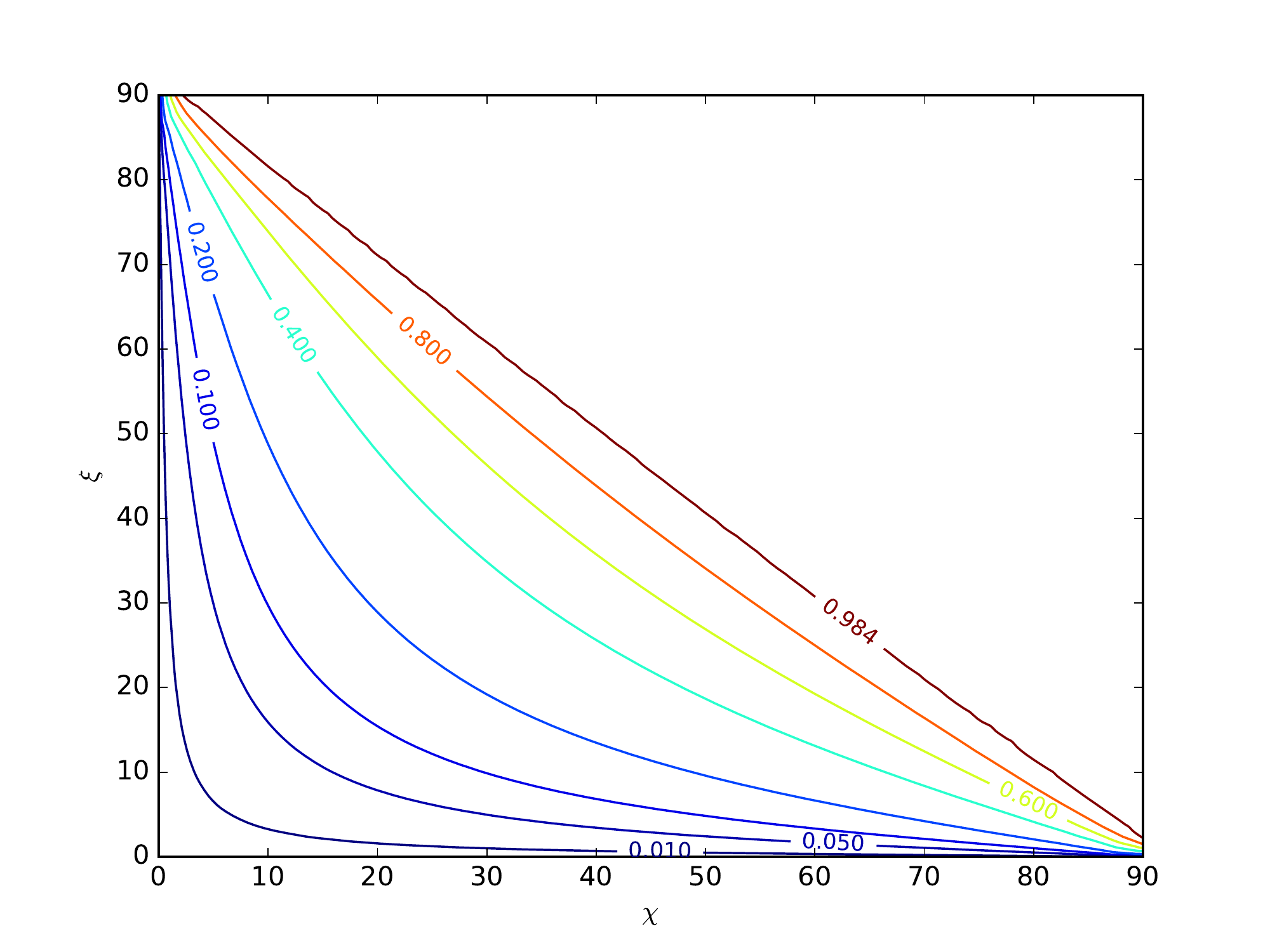}
\caption{Theoretical PF as a function of the angles $\xi$ and $\chi$, computed in this work for the source 1E 2259+586 modeled as a WD of radius $R_{\rm min}\approx 1.04\times 10^8$~cm. The left panel shows the results of the PF produced by the blackbody given by the two antipodal hot spots. The right panel shows the results for the total flux given by the blackbody plus the non-thermal power-law component, both pulsed. The observed total PF of this source in the 2--10~keV is about 20\% \citep{2008ApJ...686..520Z}.}\label{fig:PF2259}
\end{figure*}

We can measure the amount of pulsed emission by defining the \emph{pulsed fraction}
\begin{equation}
{\rm PF}=\frac{F_{\rm max}-F_{\rm min}}{F_{\rm max}+F_{\rm min}},
\end{equation}
which we show in Fig.~\ref{fig:PF2259}, as a function of the angles $\xi$ and $\chi$, for AXP 1E 2259+586. In the left panel of this figure we consider only the flux given by the blackbody produced by the two antipodal hot spots on the WD. We can see that indeed pulsed fractions as small as the above values can be obtained from magnetized WDs, for appropriate values of the geometric angles $\xi$ and $\chi$. However, the soft X-ray spectrum shows a non-thermal power-law component, additional to the blackbody one. As we have shown, the blackbody itself can contribute to the PF if produced by surface hot spots and thus the observed total PF of a source in those cases includes both contributions, mixed. It is thus of interest to explore this problem from the theoretical point of view. To do this we first recall that total intrinsic flux of this source in the 2--10~keV band is $F_{\rm tot}\approx 1.4\times 10^{-11}$~erg~cm$^{-2}$~s$^{-1}$, and the power-law flux is $F_{\rm PL} \approx 1.8 F_{\rm bb} \approx 8.5\times 10^{-12}$~erg~cm$^{-2}$~s$^{-1}$ \citep{2008ApJ...686..520Z}. The right panel of Fig.~\ref{fig:PF2259} shows the PF map for this source taking into account both the blackbody and the power-law components. By comparing this PF map with the one in the left panel which considers only the pulsed blackbody we can see that they are very similar each other. This means that in these cases where both pulsed components are in phase and have comparable fluxes it is difficult (although still possible if good data are available), to disentangle the single contributions.

%

\section{Conclusions}\label{sec:5}

We exploited the analogy with pulsars to investigate whether or not massive, highly magnetized, fast rotating WDs, can behave as neutron star pulsars and emit observable high-energy radiation. We conclude:
\begin{enumerate}
\item
We showed that WDs can produce $e^-e^+$ pairs in their magnetosphere from the decay of curvature radiation photons, i.e. we infer the structure parameters for which they are located above the WD pulsar death-line. We evaluated the rate of such a process. Then, we calculated the thermal emission produced by the polar cap heating by the pair-created particles that flow back to the WD surface due to the action of the induction electric field.

\item
In order to give a precise example of the process, we applied the theoretical results to the case of the WD model of SGRs and AXPs. We have shown that the inferred values of the WD parameters obtained from fitting with this magnetospheric emission the blackbody spectrum observed in the soft X-rays of SGRs and AXPs, are in agreement with our previous estimates using the IR, optical, and UV data, and fall within the constraints imposed by the gravitational stability of the WD.

\item
We have related the size of the spot with the size of the surface under the polar cap filled by the inward particle bombardment. We have shown that the spot area is much smaller than the polar cap area pointing to the existence of strong non-dipolar magnetic fields close to the WD surface.

\item
We have used the heat transport and energy balance equations to show that, for the actual conditions of density and temperature under the polar cap, the hot spot re-radiates efficiently the heat proportioned by the inward particle bombardment.

\item
The spot, which is aligned with the magnetic dipole moment of the WD, produces a pulsed emission in phase with the rotation period of the object. We showed that the theoretically inferred pulsed fraction of the WD span from very low values all the way to unity depending on the viewing angles. Therefore it can also account for the observed pulsed fraction in SGRs and AXPs for appropriate choices of the viewing angles. In addition, the low-energy tail of the blackbody spectrum of the hot spot could produce a non-null pulsed fraction of the flux in the optical bands as well. However, this depends on the flux produced by the surface temperature of the WD which certainly dominates the light curve at low energies. From the quantitative point of view, the size of the surface area of the spots is crucial for the explanation the observed pulsed fraction in soft X-rays. 

\item
We have also shown that the addition of a pulsed power-law component as the one observed in SGRs/AXPs does not modify appreciable the above result. The reason for this is that the non-thermal power-law component and the blackbody due to the surface hot spot have comparable fluxes and are in phase each other. In those cases it is difficult to disentangle the siingle contributions to the pusled fraction.

\end{enumerate}

We have shown that, as advanced in \citet{2013ApJ...772L..24R}, indeed the blackbody observed in the optical wavelengths of SGRs and AXPs can be due to the surface temperature of the WD, while the one observed in the X-rays can be of magnetospheric origin. For the power-law component, also observed in the soft X-rays, a deeper analysis of processes such as curvature radiation, inverse Compton scattering, as well as other emission mechanisms, is currently under study.

There is also room for application and extension of the results presented in this work to other astrophysical phenomena. WD mergers can lead to a system formed by a central massive, highly magnetized, fast rotating WD, surrounded by a Keplerian disk \citep[see][and references therein]{2013ApJ...772L..24R}. At the early stages, the WD and the disk are hot and there is ongoing accretion of the disk material onto the WD. In such a case, the WD surface shows hot regions that deviate from the spotty case, e.g. hot surface rings. That case is also of interest and will be presented elsewhere.

\section*{Acknowledgments}

It is a great pleasure to thank M.~Malheiro for thoughtful discussions and for the comments and suggestions on the presentation of our results. Likewise, we would like to thank N. Rea. We are grateful to L.~Becerra for providing us tables of the heat capacity and thermal conductivity in the range of densities and temperatures of interest for this work. D.L.C. acknowledges the financial support by the International Relativistic Astrophysics (IRAP) Ph.~D. Program. J.G.C., R.C.R.L., S. M. C. and J.A.R. acknowledge the support by the International Cooperation Program CAPES-ICRANet financed by CAPES - Brazilian Federal Agency for Support and Evaluation of Graduate Education within the Ministry of Education of Brazil. JGC acknowledges the support of FAPESP through the project 2013/15088--0 and 2013/26258--4.


\appendix

\section{Heating and cooling of particle influx bombardment}\label{app:A}

We estimate in this appendix the efficiency of the particle bombardment in heating (and re-rediating) the surface area they hit. We follow the discussion in \citet{2002ApJ...577..909G,2003A&A...407..315G} for the heat flow conditions in the polar cap surface of neutron stars, and extended it to the present case of magnetized WDs. 

The particles arriving to the surface penetrate up to a depth which can be estimated using the concept of \emph{radiation length} \citep{1980ApJ...235..576C}. For a carbon composition, the radiation length is $\Sigma \approx 43$~g~cm$^{-2}$ \citep{1974RvMP...46..815T}, so an electron would penetrate the WD surface up to a depth
\begin{equation}\label{eq:depth}
\Delta z\approx \frac{\Sigma}{\rho} = 4.3\times 10^{-3}\,{\rm cm}\left(\frac{10^4\,{\rm g\,cm}^{-3}}{\rho}\right).
\end{equation}
With the knowledge of the thickness of the layer under the surface where the energy deposition occurs we can proceed to estimate the properties of the diffusion and re-radiation of the kinetic energy of the particle influx using the heat transport and energy balance equations on the star's surface corresponding to the polar cap. The typically small distances [see Eq.~(\ref{eq:depth})] allow us to introduce a plane-parallel approximation in the direction parallel to the magnetic field lines, say in the direction $z$ orthogonal to the surface. 

The energy balance can be simply written as
\begin{equation}\label{eq:balance}
F_{\rm rad} = F_{\rm heat} + F_{\rm cond},
\end{equation}
where $F_{\rm heat} = e \Delta V \eta \rho_{\rm GJ} c$, $F_{\rm cond}=-\kappa \partial T/\partial z$ and $F_{\rm rad} = \sigma T^4$, with $\kappa$ the thermal conductivity (along the $z$-direction).  

Let us first estimate the characteristic cooling time. To do this we switch off energy losses and heating terms in the energy balance equation (\ref{eq:balance}), i.e. the radiation flux is only given by conduction:
\begin{equation}\label{eq:conduction}
\sigma T^4 =-\kappa \frac{\partial T}{\partial z},
\end{equation}
which leads to the heat transport equation
\begin{equation}\label{eq:Htransport}
c_v \frac{\partial T}{\partial t} = \frac{\partial}{\partial z}\left( \kappa \frac{\partial T}{\partial z} \right).
\end{equation}
where $c_v$ is the heat capacity per unit volume. We can therefore obtain the characteristic ($e$-folding) cooling and heating time assuming the quantities are uniform within the penetration depth $\Delta z$, i.e. 
\begin{equation}\label{eq:tcoolheat}
\Delta t_{\rm cool} = \frac{\Delta z^2 c_v}{\kappa},\qquad \Delta t_{\rm heat} = \frac{c_v \Delta z}{\sigma T^3}.
\end{equation}
We can now introduce the radiation to heating efficiency parameter 
\begin{equation}\label{eq:epsilon}
\epsilon \equiv \frac{F_{\rm rad}}{F_{\rm heat}} = \frac{1}{1+\Delta t_{\rm heat}/\Delta t_{\rm cool}} = \frac{1}{1+\kappa/(\sigma T^3 \Delta z)},
\end{equation}
which shows that in equilibrium, $\Delta t_{\rm heat} = \Delta t_{\rm cool}$, we have $\epsilon = 1/2$. 

In estimating the spot temperature (\ref{eq:TspotIG}) we have assumed in Eq.~(\ref{eq:Tspot1}) full re-radiation of the influx, namely we assumed $\epsilon = 1$. We proceed now to estimate the realistic values of $\epsilon$ from Eq.~(\ref{eq:epsilon}) to check our assumption. We compute the thermal conductivity from \citet{1993ApJ...418..405I} and the heat capacity from \citet{1998PhRvE..58.4941C,2000PhRvE..62.8554P}. For example, at a density $\rho=10^3$~g~cm$^{-3}$ and $T=10^6$~K, we have $c_v=2.7\times 10^{10}$~erg~cm$^{-3}$~K$^{-1}$ and $\kappa\approx 4\times 10^{11}$~erg~cm$^{-1}$~s$^{-1}$~K$^{-1}$, and Eq.~(\ref{eq:epsilon}) gives $\epsilon \approx 0.86$. At $T=10^7$~K, we have $c_v=3.8\times 10^{11}$~erg~cm$^{-3}$~K$^{-1}$ and $\kappa\approx 3.4\times 10^{13}$~erg~cm$^{-1}$~s$^{-1}$~K$^{-1}$ and $\epsilon \approx 1$. 

\bsp
\label{lastpage}
\end{document}